# Strain-Tuned Magnetic Anisotropy in Sputtered Thulium Iron Garnet Ultrathin Films and TIG/Au/TIG Valve Structures

Gilvânia Vilela[1,2], Hang Chi[1,3], Gregory Stephen[4], Charles Settens[5], Preston Zhou[6], Yunbo Ou[1], Dhavala Suri[1], Don Heiman[1,4] and Jagadeesh S. Moodera[1,7]

[1] Plasma Science and Fusion Center, and Francis Bitter Magnet Laboratory, Massachusetts Institute of Technology, Cambridge, Massachusetts 02139, USA
[2] Física de Materiais, Escola Politécnica de Pernambuco, UPE, Recife, Pernambuco 50720-001, Brasil
[3] U.S. Army CCDC Army Research Laboratory, Adelphi, Maryland 20783, USA
[4] Department of Physics, Northeastern University, Boston, Massachusetts 202115, USA
[5] Materials Research Laboratory, Massachusetts Institute of Technology, Cambridge, Massachusetts 02139, USA
[6] Department of Electrical Engineering, California Institute of Technology, Pasadena, California 91125, USA
[7] Department of Physics, Massachusetts Institute of Technology, Cambridge, Massachusetts 02139, USA

Electronic mail: moodera@mit.edu, gilvania.vilela@upe.br

**Abstract**

Defining the magnetic anisotropy for in-plane or out-of-plane easy axis in ferrimagnetic insulators films by controlling the strain, while maintaining high-quality surfaces, is desirable for spintronic and magnonic applications. We investigate ways to tune the anisotropy of amorphous sputtered ultrathin thulium iron garnet (TIG) films, and thus tailor their magnetic properties by the thickness (7.5 to 60 nm), substrate choice (GGG and SGGG), and crystallization process. We correlate morphological and structural properties with the magnetic anisotropy of post-growth annealed films. 30 nm thick films annealed at 600 °C show compressive strain favoring an in-plane magnetic anisotropy (IPMA), whereas films annealed above 800 °C are under a tensile strain leading to a perpendicular magnetic anisotropy (PMA). Air-annealed films present a high degree of crystallinity and magnetization saturation close to the bulk value. These results lead to successful fabrication of trilayers TIG/Au/TIG, with coupling between the TIG layers depending on Au thickness. These results will facilitate the use of TIG to create various in situ clean hybrid structures for fundamental interface exchange studies, and towards the development of complex devices. Moreover, the sputtering technique is advantageous as it can be easily scaled up for industrial applications.

## 1. Introduction

There has been widespread activity and interest in employing ferromagnetic insulator (FI) materials for developing spintronics, caloritronics, and topotronics devices [1-13]. The magnetic state of the FI layers can be manipulated by pure spin currents, which reduce the Joule heating and the need for an external magnetic field. When FI materials are in close contact with heavy metals, topological insulators (TIs), superconductors (SCs), graphene and two-dimensional (2D) materials, the strongly correlated interaction from the proximitized interface results in an effective local magnetic field of tens to hundreds of tesla [14-16]. This comes from the electrostatic and the exchange interaction with magnetic ions at the interface. Consequently, subjected to such enormous exchange fields, the adjacent FI layer greatly modulates these quantum materials. This interfacial exchange coupling between FI and these new materials creates a very rich interface yet to be well understood and explored, and whose success depends on the quality of the interface, magnetic properties and crystallinity of each layer.

Rare-earth iron garnets (RIGs) are good FI candidates for advanced applications because of their high Curie temperature ($T_c$), high electrical resistivity, and low magnetic losses even at high frequencies [17-30]. Among RIGs, yttrium iron garnet (YIG – $Y_3Fe_5O_{12}$) and thulium iron garnet (TIG - $Tm_3Fe_5O_{12}$) are very suitable for spin transport studies and applications, because they don't present a compensation temperature, due to their higher number of electrons in the $4f$ shell of Y and Tm [31-33].

The optical, electronic, magnetic, and lattice-dynamical properties of films depend on strain present in the films. Our goal is to study the dependence and control of the magnetic anisotropy in TIG thin films: their easy-axis directions depend on the shape anisotropy, magnetostriction, and in-plane strain. For materials with a negative magnetostriction constant ($\lambda < 0$), as in YIG and TIG, this contribution favors an out-of-plane easy axis for films under tension, whereas it favors an in-plane easy axis for films under compression [34-37]. TIG with (111) orientation has a magnetostriction constant of $\lambda_{111} = -5.2 \times 10^{-6}$, almost twice the value for YIG [38], and thus makes it easier to have a perpendicular magnetic anisotropy (PMA) in TIG films. An important application of PMA is the increase in information density in hard disk drives [39]. Going forward, PMA will be crucial for the study of new physics phenomena, such as the quantum anomalous Hall effect in exchange-coupled FI/TI structures, where the PMA is essential for breaking the time-reversal symmetry in TIs [40, 41].

TIG thin films fabricated by pulsed laser deposition (PLD) are successfully employed in heterostructures and devices combined with other materials such as heavy metals and TIs [11, 12, 36, 42]. These films are usually grown at high temperatures (~ 800 °C) in an oxygen atmosphere on gadolinium gallium garnet (GGG) or substituted GGG (SGGG) substrates and show a saturation magnetization of ~ 100 emu/cm³, with root mean square (rms) roughness of ~ 0.15 nm. Although PLD is a promising technique for growing complex compound films as it easily transfers the material from the target to the substrate, this technique limits the area of growth and hence large-scale fabrication. There are other issues as the particulate incorporation in ablated films and non-uniform deposition, which compromise the manufacturability of multilayer structures [43, 44].

Recently, there have been reports of off-axis sputtered TIG films, where the films are grown at room temperature in a mixture of oxygen and argon, followed by post-growth annealing. The films are epitaxial with low coercive fields, rms roughness of 0.2 nm and saturation magnetization of 99 emu/cm³ [45, 46]. The sputtering technique is suitable for growing uniform films and *in situ* heterostructures over large substrate areas. However, there is little information detailing how the growth and annealing conditions affect the relationship between structure and magnetic properties in TIG thin films.

In this study, we observe that it is possible to tune the magnetic anisotropy by adjusting the film thickness, the annealing parameters such as temperature, atmosphere, anneal duration, and choice of substrate. Our results show how to switch the easy axis from out-of-plane to in-plane, for example, by lowering the annealing temperature for a 30 nm thick TIG film. We also observe that air-annealed films are very smooth and have well-defined PMA, which reduces the cost of fabricating high-quality epitaxial TIG films. As a result of this study, we are able to fabricate epitaxial trilayers composing of two TIG layers separated by a thin Au film, TIG/Au/TIG. The strength of the coupling between the TIG layers depends on the thickness of the Au film. Future studies on these trilayers should show that a pure spin current flowing from the first to the second TIG layer results in a torque on the second layer when the magnetizations are misaligned [47].

## 2. Thin Film Fabrication and Morphological Characterization

TIG thin films were grown on (111)-oriented crystalline substrates of GGG ($Gd_3Ga_5O_{12}$) and SGGG [$(Gd_{2.6}Ca_{0.4})(Ga_{4.1}Mg_{0.25}Zr_{0.65})O_{12}$]. Before deposition, the substrates were annealed to improve their surface crystallinity, using a quartz tube furnace at 1000 °C for 6 hours, in 1 atm oxygen pressure. After this procedure, the substrates presented atomically flat terraces, as can be seen in the 5 $\mu m \times 5 \mu m$ atomic force microscopy (AFM) scan image shown in Fig. 1(a). The rms roughness within any terrace is 0.10 nm, and the step's average height between the terraces is 0.18 nm, which corresponds to the (444)-GGG reciprocal lattice node.

After the thermal treatment, the substrates were placed on a Mo holder inside a load lock chamber connected to an ultra-high vacuum (UHV) sputtering chamber. Inside the sputtering chamber, with base pressure below $5 \times 10^{-8}$ Torr, the substrates were degased at 300°C for one hour and then cooled to room temperature for the growth of TIG films. A 99.9% pure TIG target was sputtered at a rate of 1.4 nm/min, with radio frequency (RF) power of 50 W and a working argon pressure of 2.8 mTorr. To improve the uniformity of TIG films, the substrate holder was rotated during film deposition. The angle between the substrate holder's plane and the sputtering gun was 30°, while the distance between them was 20 cm.

It is important to note that different substrate temperatures were used during the growth of TIG in an attempt to obtain as-grown epitaxial sputtered films with non-negligible magnetic moments. While maintaining the same growth parameters with pure Ar working pressure of 2.8 mTorr and 50 W of RF power, the TIG films grown at room temperature, 350 °C or 700 °C displayed magnetic signals only after post-growth thermal annealing. We also sputtered TIG films in a mixture of argon and oxygen, with the oxygen partial pressure of 2% of 2.8 mTorr, at both room temperature and 700 °C. Again, as-grown TIG films didn't show magnetic signal. For this reason, we focus this study on TIG films sputtered in a pure Ar at room temperature.

When TIG films are under tension ($\sigma > 0$) the easy axis tends to be normal (perpendicular) to the film's plane. If it is under

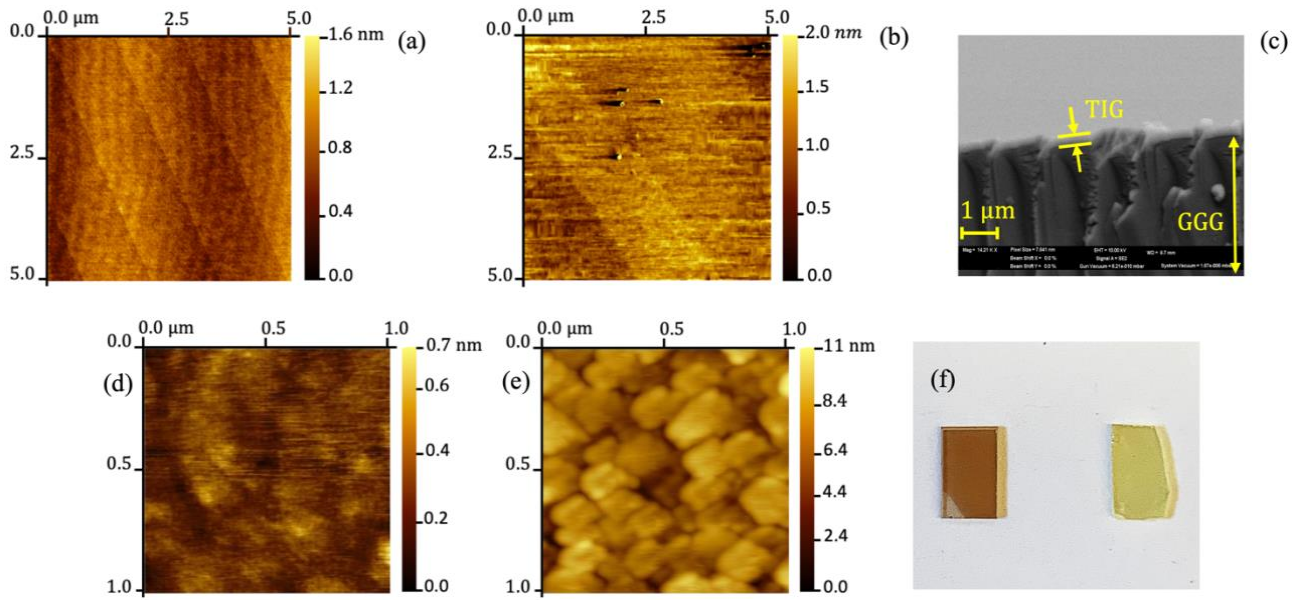

Fig. 1: Surface morphologies of the substrate and the sputtered TIG thin films. (a) 5 $\mu$m × 5 $\mu$m AFM scan of 1000 ℃-annealed (111)-oriented GGG. The scan shows atomically flat terraces with rms roughness of 0.11 nm inside the monoatomic steps. (b) 5 $\mu$m × 5 $\mu$m AFM scan of as-grown 30 nm thick TIG film on GGG. The rms roughness is 0.20 nm. (c) Cross-sectional SEM image of 200 nm thick TIG film over GGG. (d) and (e) show 1 $\mu$m × 1 $\mu$m AFM scans of 800 ℃-annealed 30 nm thick TIG films on GGG and SGGG, respectively. TIG/GGG is smoother, with rms roughness below 0.10 nm, while TIG/SGGG has a rms roughness of 0.60 nm. (f) Pictures of 30 nm thick as-grown (brown color) and 800 ℃-annealed (smooth yellow) TIG films over 2 mm × 4 mm GGG substrates.

Figures 1 (b) and (c) show AFM and scanning electron microscopy (SEM) scans of as-grown films on GGG with thicknesses of 30 and 200 nm, respectively. These scans indicate a smooth and uniform surface with rms roughness of 0.20 nm. To crystallize TIG films, they were annealed under different conditions of temperature, atmosphere, and duration. Figures 1(d) and (e) show 1 $\mu$m × 1 $\mu$m AFM scans of 30 nm thick post-annealed TIG films on GGG and SGGG, respectively. These films were annealed for 8 hours at 800 ℃ in flowing oxygen. TIG/GGG displays a rms roughness of 0.10 nm, smoother than that of TIG/SGGG (0.60 nm). The as-grown films were brown and moderately transparent, whereas, after annealing, they became highly transparent with yellow color (see Fig. 1(f)). In the next section, we discuss in detail how the film thickness and the annealing conditions affect the crystal and magnetic properties of TIG thin films.

## 3. Strain-Dependent Magnetic Properties

Thermal expansion and lattice mismatch between an epitaxial film and a substrate are expected to generate mechanical stress, $\sigma$. Besides, if the film is magnetostrictive, this stress creates a uniaxial magnetic anisotropy that manifests itself by creating anisotropy in the film [34, 48].

compression ($\sigma < 0$), the easy axis tends to lie in-plane.

It is thus possible to tune the magnetic anisotropy of TIG films by controlling the strain due to lattice mismatch and annealing conditions. When the perpendicular anisotropy field, $H_\perp$, is strong enough to overcome the shape anisotropy ($H_\perp > 4\pi M_S$), the film presents a normal easy axis. Along (111) orientations, $H_\perp$ is defined as [35]:

$$H_\perp = \frac{-4K_1 - 9\lambda_{111}\sigma}{3M_S}, \quad (1)$$

where $K_1$ is the first-order cubic anisotropy constant (with the second-order $K_2$ assumed to be zero), $\lambda_{111}$ is the magnetostriction constant, $\sigma$ is the in-plane stress and $M_S$ is the saturation magnetization. The stress in the film, $\sigma$, is a function of the elastic stiffness constants of the deformation tensor and is directly proportional to the in-plane strain ($\varepsilon_\parallel$) [49]. We estimate the in-plane strain for the sputtered TIG thin films as the percent variation from their in-plane lattice constant, $a$, as compared to the reported value for unstrained samples $a_{TIG} \approx 1.233$ nm [31, 32]:

$$\varepsilon_\parallel = \frac{a - a_{TIG}}{a_{TIG}} \times 100\ \%. \quad (2)$$

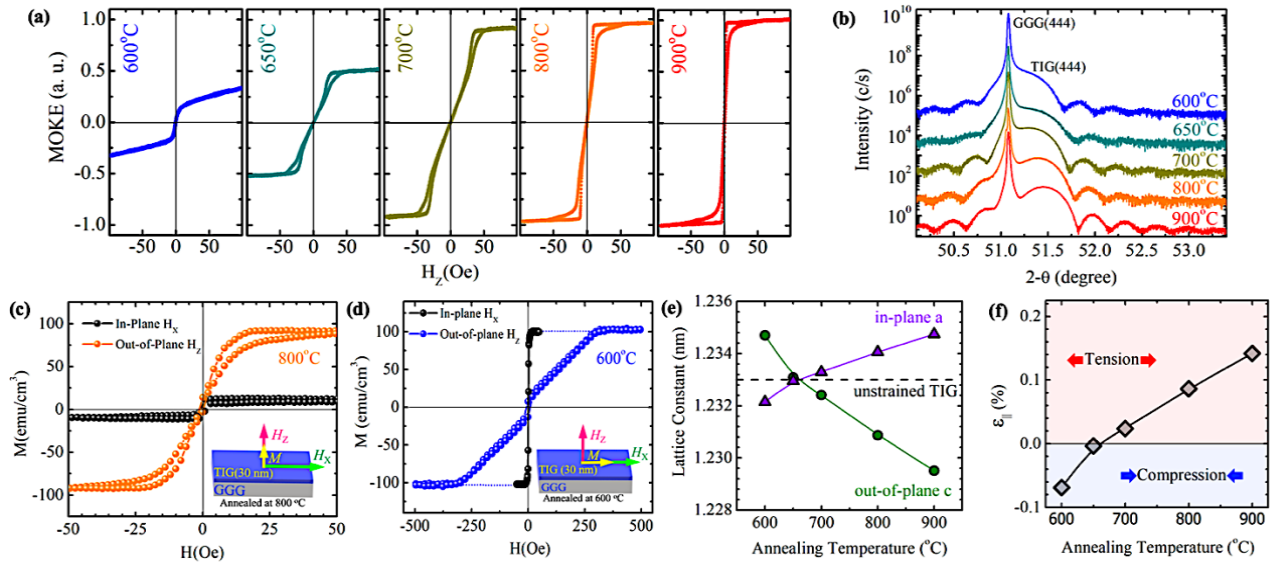

Fig. 2: Polar MOKE, SQUID and X-ray diffraction for 30 nm thick TIG films annealed at different temperatures, for 8 hours in an oxygen atmosphere. (a) and (b) Show polar MOKE signals and XRD patterns for films annealed at $T_a$, with 600 °C < $T_a$ < 900 °C. (c) And (d) show SQUID measurements for films annealed at 800 °C and 600 °C, respectively, with field applied in-plane ($H_X$) and out-of-plane ($H_Z$) of the film, as schematized in the insets. The film annealed at 800 °C shows a PMA, whereas the film annealed at 600 °C shows an IPMA. Both samples presented magnetization saturation of $M_S = 100$ emu/cm³. (e) In-plane, $a$, and out-of-plane, $c$, lattice constants as a function of the annealing temperature. (f) In-plane strain, $\varepsilon_\parallel$, versus annealing temperature. $\varepsilon_\parallel$ is negative for 600 °C and the film is under compression, which favors an IPMA. Above 650 °C, the in-plane strain is positive, as the film is under a tension which favors a PMA. Films annealed at 650 °C are relaxed, $\varepsilon_\parallel \approx 0$. Error bars are smaller than the data point symbol. All the solid lines are guides for eyes.

The out-of-plane, $c$, and in-plane, $a$, lattice constants for the TIG films are extracted from high-resolution X-ray diffraction (XRD) patterns, where $c$ is the lattice spacing of the cubic crystal calculated using Bragg's law. The in-plane lattice constant $a$ is estimated considering that the volume of the TIG unit cell remains constant under the tetragonal distortion, $v = (1.233nm)^3 = a^2 \times c$, thus $a \approx \sqrt{(1.233nm)^3/c}$.

*3.1 Influence of annealing temperature on strain and magnetic properties*

We investigate how the choice of the annealing temperature ($T_a$) affects the in-plane strain in the 30 nm thick TIG films, therefore controlling the magnetic behavior. For this purpose, a set of 30 nm thick TIG films were annealed at different temperatures of 600 °C, 650 °C, 700 °C, 800 °C and 900 °C, for a duration of 8 hours in flowing oxygen. The films were all deposited at the same time, on the same batch of GGG substrates to guarantee that $T_a$ is the only parameter differentiating the behavior of the samples. The temperature of the quartz tube furnace with the film inside was increased at a rate of 10 °C/min until it reached $T_a$, and then remained at $T_a$ for 8 hours. After the high temperature annealing, the furnace was shut off to cool down naturally in the presence of oxygen to about 150 °C (~ 3-4 hr) before removing the films.

As $T_a$ increases, the out-of-plane magnetization component increases, showing sharp switching at smaller coercive fields, as seen in the magneto-optic Kerr effect (MOKE) results in Fig. 2(a). The laser used in the MOKE setup has a wavelength of 635 nm and a power o 4.5 mW. As the perpendicular magnetic field ($H_Z$) increases up to 75 Oe, we observe the saturation in the out-of-plane direction of the magnetic moments, for 30 nm thick samples annealed at temperatures above 700 °C. On the other hand, samples annealed below 650 °C don't achieve the saturated state up to a field of 75 Oe. Stronger fields are required to align all the magnetic moments in the perpendicular direction. The curved shapes observed in the hysteresis loops might be related with labyrinth domain structures very common in garnet films [50-53].

Starting from a saturated magnetized state with a positive $H_Z$, as the field decreases, the magnetization reversal process starts at a perpendicular reversal field, $H_R$, that is positive for 30 nm thick TIG films. $H_R$ is 3.7 Oe for the film annealed at 900 °C, which increases to 317 Oe for the film annealed at 600 °C. In the latter case, $H_R$ is extracted from the superconducting quantum interference device (SQUID) measurements that clearly show a switch of the easy axis from out-of-plane, for films annealed at 800 °C in Fig. 2(c), to in-plane for films annealed at 600 °C in Fig. 2(d). After substracting the paramagnetic background from the GGG substrate, the saturated magnetization for both films is estimated to be 100 emu/cm³, close to the reported value of 110 emu/cm³ for bulk TIG [38].

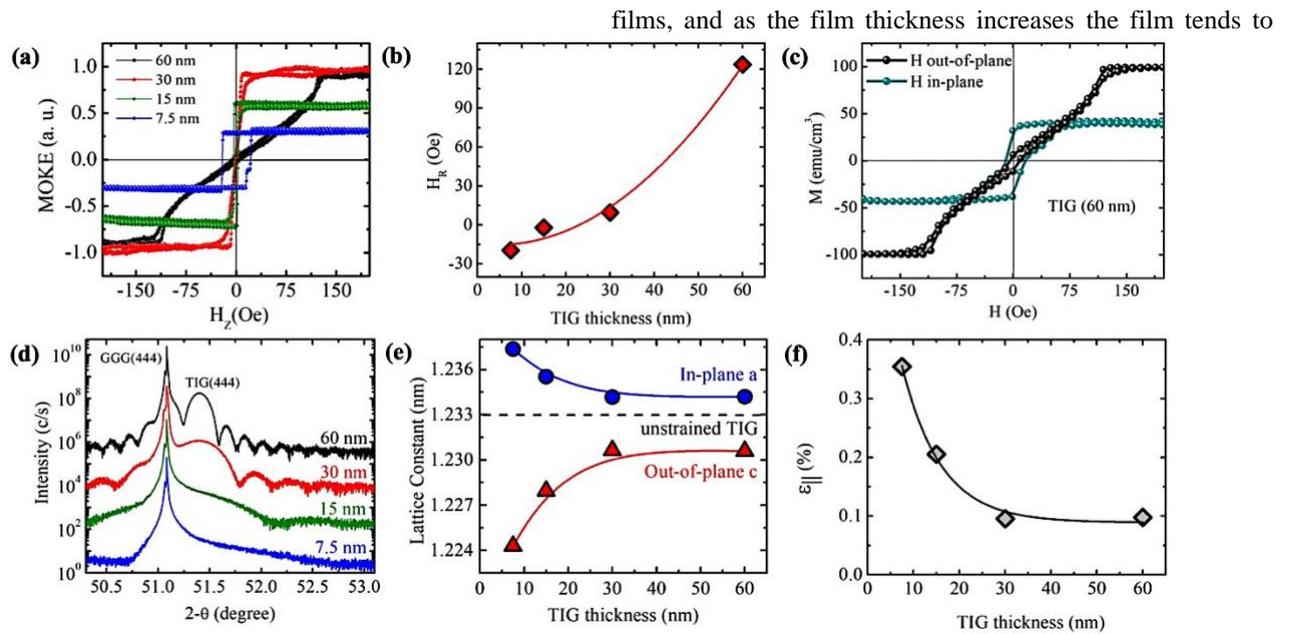

Fig. 3: The thickness (*d*) dependence of magnetization and in-plane strain of TIG films. (a) Polar MOKE for 7.5 nm < *d* < 60 nm thick TIG films, annealed for 8 hours in oxygen. The thinnest film shows a square hysteresis loop, a signature of a strong PMA. (b) Reversal perpendicular magnetic field, $H_R$, versus film thickness, shows a non-linear increase as *d* increases. (c) Magnetic hysteresis loop measured with a SQUID magnetometer for a 60 nm thick TIG film annealed at 800 °C for in-plane and out-of-plane applied fields. (d) XRD patterns for four TIG films with different thickness. (e) XRD data analysis shows an increase of the in-plane lattice constant, *a*, as the film thickness decreases. (f) In-plane strain versus TIG film thickness. Thinner films are under higher tensile in-plane strains, which favored a robust PMA. Error bars are smaller than the data point symbol. All the solid curves are guides for eyes.

To understand the role of the in-plane strain $\varepsilon_\parallel$ in the magnetic anisotropy of TIG films, high-resolution XRD patterns were measured for each sample, see Fig. 2(b). As $T_a$ increases from 600 °C to 900 °C, the 2θ diffraction peak corresponding to the (444) plane moves from 51.21° to 51.40° due to an increase of the out-of-plane lattice constant *c*, accompanied by a decrease of the in-plane lattice constant *a*, as shown in Fig. 2(e). Alternatively, $\varepsilon_\parallel$ is negative - 0.070 ± 0.005 % for films annealed at 600 °C, therefore, the film is under a compressive strain and thus favoring an in-plane easy axis. Upon increasing $T_a$, the strain increases monotonically to + 0.142 ± 0.005 % for 900 °C anneal, as shown in Fig. 2(f).

Overall, these results show that 30 nm thick TIG films are unstrained $\varepsilon_\parallel \approx 0$ for annealing at 650 °C. For $T_a = 600$ °C the films are under a compressive strain and present an IPMA, and for $T_a > 650$ °C the films are under a tensile strain that overcomes the shape anisotropy and develops a PMA as $T_a$ approaches to 900 °C.

*3.2 Strain-induced magnetic properties versus TIG film thickness*

The thickness is also an important parameter to control the magnetic anisotropy in TIG films. The strain in epitaxial films is significantly influenced by the substrate lattice for thinner films, and as the film thickness increases the film tends to relax. Various films with different thickness, 7.5, 15, 30 and 60 nm, were fabricated on (111)-GGG substrates, under the same sputtering and annealing conditions. After growth, the films were annealed together at 800 °C for 8 hours in oxygen flow. Figure 3(a) shows the polar MOKE signals for each one of the samples. As seen for the thinnest film, the hysteresis loop is square and the magnetization reversal occurs abruptly in a negative field $H_R$ = - 20 Oe upon reversing the applied field. As the film becomes thicker, $H_R$ increases non-linearly towards positive values, as shown in Fig. 3(b). SQUID magnetometry for the 60 nm thick TIG film yields magnetization values of 42.3 emu/cm³ for the in-plane component, and 99.2 emu/cm³ for the out-of-plane, as shown in Fig. 3(c), in an applied field of 200 Oe.

The XRD scan analysis reveals an enhancement of $\varepsilon_\parallel$ as the film becomes thinner, see Figs. 3(d), (e) and (f). Both *a* and $\varepsilon_\parallel$ increase as film thickness decreases. For thicker films, above 30 nm, the films appear to relax and these parameters do not lead to appreciable changes.

*3.3 Strain-induced magnetic properties versus annealing duration, atmosphere and substrate lattice*

In addition to the role played by the annealing temperature and film thickness, we also investigate the influence of atmosphere and duration for annealing, as well as the substrate lattice constant. For this study, the 30 nm thick TIG films, from the same batch, were annealed in flowing gases of $O_2$, Ar, air, a mixture of 95% Ar and 5% $O_2$ and vacuum, respectively. $T_a$ was kept at 800 ℃ and the duration was 8 hours. Except in vacuum, high-quality TIG films were obtained in all annealing environments. While the polar MOKE results, Fig. 4(a), show no significant changes for different annealing atmospheres, the magnitude of $H_R$ increases as the $O_2$ content in the annealing atmosphere decreases, as seen in Fig. 4(b). The amount of oxygen present during the annealing process might affect the film's surface roughness, which affects their domain structure, and so affects $H_R$ [54, 55]. The maximum $\varepsilon_\parallel$ is achieved for samples annealed in $O_2$, while it is minimum for Ar annealed samples. All samples, except that annealed in vacuum, develop positive strain and intense perpendicular magnetic signals, as shown in Fig. 4(d). Having oxygen present during the annealing process prevents the TIG films from losing oxygen. Moreover, the annealing pressure could also be an important parameter which wasn't explored in this study.

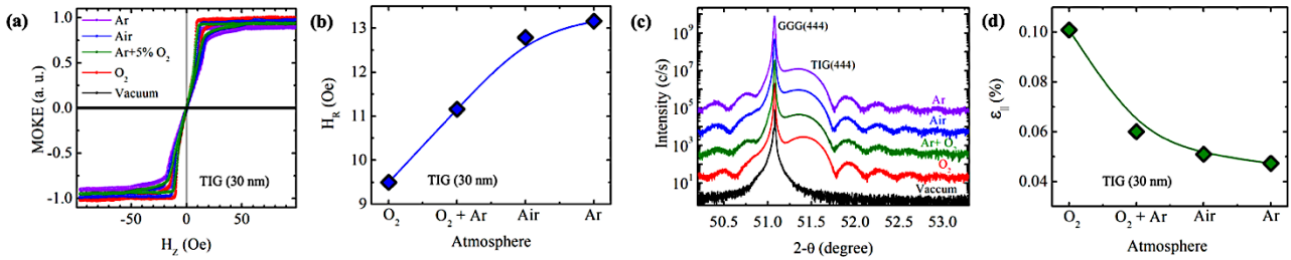

Fig. 4: Dependence of polar MOKE and in-plane strain with the annealing atmosphere for 30 nm thick TIG films. (a) Polar MOKE signals for annealing under pressures of oxygen, argon, air, 95% argon + 5% oxygen and vacuum, for 8 hours at 800 ℃. (b) The analysis of the hysteresis loops shows an increase of $H_R$ as the amount of oxygen present in the atmosphere decreases. (c) XRD scans as a function of the atmosphere show peaks for TIG with Laue fringes, indicating high crystallinity, with the exception for vacuum. (d) The in-plane strain is positive for all samples and increases as the amount of oxygen in the annealing atmosphere increases. Error bars are smaller than the data point symbol. All the solid curves are guides for eyes.

To understand the influence of the annealing duration, 30 nm thick TIG films, from the same batch, were annealed for 30 minutes and 8 hours. TIG film crystallizes and displays a strong perpendicular magnetic signal within 30 minutes of annealing, as shown in figures 5(a) and (b). This short annealing was carried out in an oxygen atmosphere at 800 ℃.

The value of $\varepsilon_\parallel$ only doubles from 0.061 ± 0.005 % to 0.105 ± 0.005 % by increasing the annealing duration from 30 minutes to 8 hours, showing that major structural changes happen quickly during annealing.

To study the substrate influence, 30 nm of TIG were sputtered on (111)-oriented SGGG and GGG substrates in the same run. After growth, the samples were annealed together at 800 ℃ for 8 hours in $O_2$ atmosphere. The lattice constant of SGGG is $a_S$ = 1.2480 nm, which is 0.84% higher than that for GGG ($a_S$ = 1.2376 nm), calculated using Bragg's law for (444) peaks diffraction measured at 50.43° for SGGG, and at 51.07° for GGG. Figures 5(c) and (d) show how the magnetic anisotropy strongly depends on the lattice mismatch between the film and the substrate. Even a small change of 0.84%, drastically modifies the polar MOKE and XRD results. The XRD scans for TIG/SGGG don't show Laue fringes on either side of the film peak, which indicates the film is not as smooth as the TIG film on GGG. This is confirmed by the AFM scans shown in Figs. 1(d) and (e). The coercive field for the TIG film on GGG is 5.3 Oe, whereas for the film on SGGG is 276 Oe, an enhancement of more than 5000 %. The value of $\varepsilon_\parallel$ changes

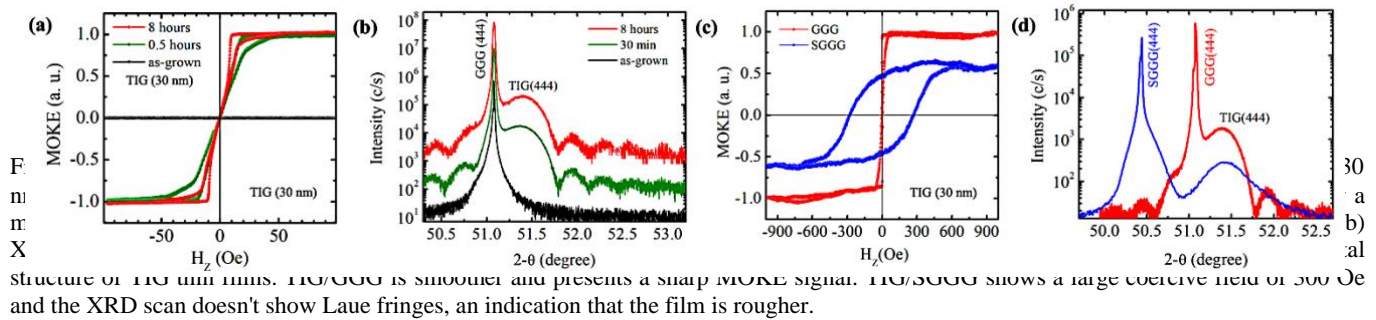

Fig. 5: [partially obscured] ... structure of TIG thin films. TIG/GGG is smoother and presents a sharp MOKE signal. TIG/SGGG shows a large coercive field of 300 Oe and the XRD scan doesn't show Laue fringes, an indication that the film is rougher.

from $0.092 \pm 0.005$ % for GGG to $0.124 \pm 0.005$ % for SGGG substrates.

Finally, Fig. 6 summarises the overall film properties on growth and annealing conditions: how the annealing conditions, film's thickness and substrate's lattice constant can be used to build up or to strengthen a PMA in sputtered TIG thin films deposited over GGG(111). As the strain increases, indicated by the green arrow in Fig. 6, the PMA emerges or becomes stronger. The conditions to strengthen the PMA are: high annealing temperatures, thinner films, annealing in oxygen enriched atmosphere for longer duration, and higher lattice mismatch between the film and the substrate.

*3.4 TIG/Au/TIG epitaxial trilayers*

The results obtained in this work show that it is possible to select a set of parameters for thickness, growth and annealing conditions to control the magnetic anisotropy of TIG films. In principle, it is possible to fabricate multilayers of TIG combined with other materials, with each TIG layer presenting a selected PMA or IPMA. Here we focus in the fabrication process of epitaxial trilayers TIG/Au/TIG, as schematized in Fig. 7(a), for future applications in magnetoresistive devices and magnon valves.

The ability to build devices composed by magnetic films with PMA has advantages for spintronic applications, such as higher device density similar to that used in present PMA recording media [56]. Magnetic films with PMA are preferable for patterned devices because they present more uniform magnetization, and they are less susceptible to thermal instabilities due to magnetization curling at the edges when compared with in-plane magnetic materials [57]. Moreover, it is easier to reverse an out-of-plane magnetization by spin transfer torque (STT) than an in-plane magnetization, and less switching current is required in the first case [58,59]. For these reasons, we investigate how to tune the PMA in TIG/Au/TIG. In this case, it is required thinner layers of TIG combined with higher annealing temperatures.

It is also possible to explore the IPMA in TIG/Au/TIG for developing magnon valves, where Au is a heavy metal (HM) with strong spin-orbit coupling [60]. As discussed previously, a 30 nm thick TIG film annealed at $600\ °C$ has an in-plane easy axis. The net pure spin current, generated by means of a longitudinal spin Seebeck effect, for example, flowing through TIG/Au/TIG depends on the relative orientation between the two magnetic layers [9, 47]. When the spin polarization $\hat{\sigma}$ and the spin current $\vec{J}_S$ are not parallel, it is possible to detect a non-zero charge current $\vec{J}_C$ in a Pt Hall bar placed on the top of TIG/Au/TIG, by means of the inverse spin

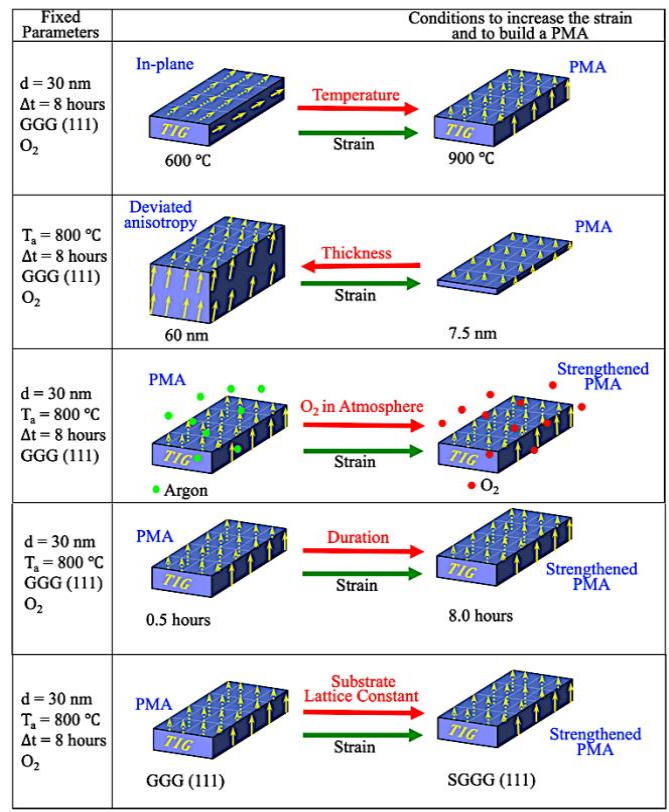

Fig. 6: Summary of how the annealing parameters, film's thickness and substrate's lattice constant build up or strengthen the PMA in sputtered TIG thin films. The left column shows a set of fixed parameters, while a variable parameter is analysed. The variable parameter is indicated by the red arrow. The green arrow indicates the increase of the strain, and thus, indicating the way to achieve higher PMA.

Hall effect (ISHE) expressed as $\vec{J}_C = \theta_H\ \hat{\sigma} \times \vec{J}_S$ [61-64], where $\theta_H$ is the Hall angle. Here the excited spin current $\vec{J}_S$ has the same direction of the thermal gradient which is perpendicular to the sample's plane, while $\hat{\sigma}$ has the same orientation of the magnetization. Magnon valves YIG/Au/YIG and YIG/NiO/YIG have been successfully fabricated [65, 66]. For favoring an in-plane magnetic anisotropy it is required thicker TIG films with lower annealing temperatures.

The trilayers TIG/Au/TIG, were grown over annealed (111)-GGG substrates. TIG films were sputter deposited, while Au films were deposited in a thermal evaporator. Subsequently, the entire structure was subjected to a post-growth annealing at $700\ °C$ for 2 hours in flowing oxygen. It turned out that after the annealing, all three layers are epitaxial, and rather remarkably, the TIG layer on top of the Au layer also displays PMA. The epitaxial growth of Au (111) over the bottom TIG layer is possible because the lattice constant of Au (0.406 nm) is approximately one-third of TIG's lattice constant (1.233 nm). The analysis of the XRD data for TIG(30 nm)/Au(8 nm)/TIG(15 nm) shows one peak for Au, and two superimposed peaks for the bottom TIG(30 nm) and for the top TIG (15 nm), as indicated in Fig. 7(b). The strain calculated for the bottom and top TIG layers are + 0.168 ± 0.005 %, and + 0.510 ± 0.005 %, respectively. These values are greater than the strain for simple films of TIG deposited over GGG(111), therefore these strains are strong enough to build up a PMA in each TIG layer. However, when TIG was grown on top of Au films that were deposited on (0001)-sapphire substrates, the epitaxy was unsuccessful: XRD scans showed only the (111) peak of Au, and the magnetic signal for the TIG was negligible.

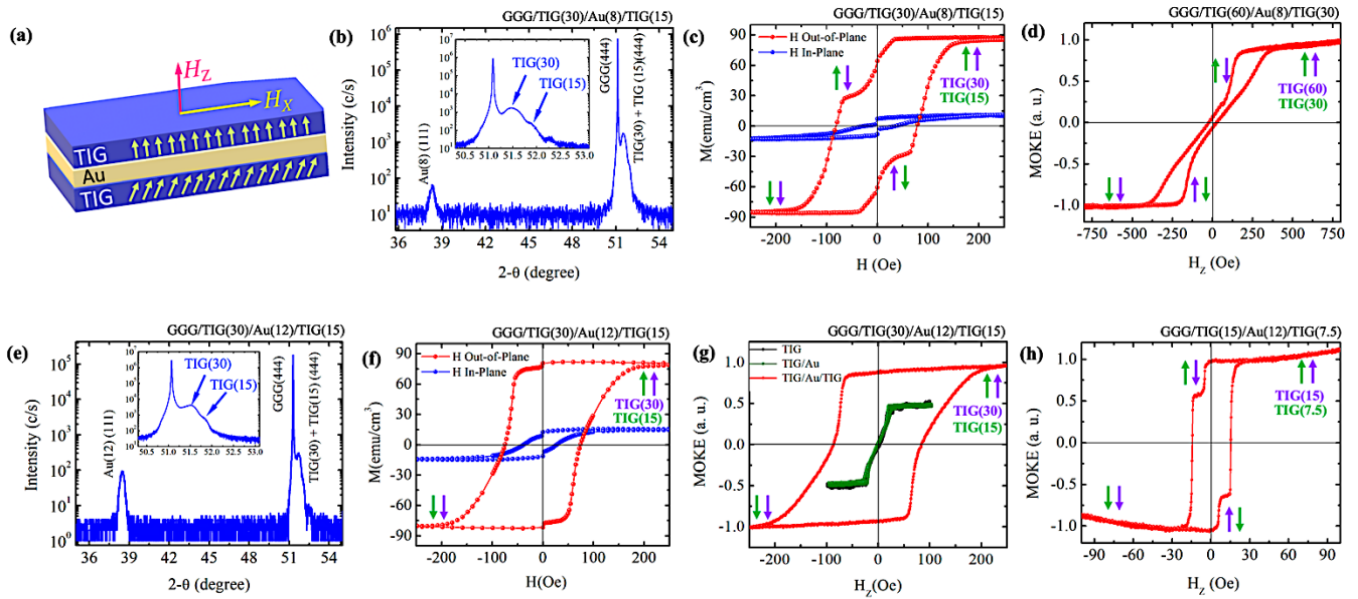

Fig. 7: Magnetic and structural characterization for TIG/Au/TIG over GGG substrate. (a) Schematic figure of the trilayer and the in-plane, $H_X$, and out-of-plane, $H_Z$, applied magnetic field configurations. (b) XRD scan shows diffraction peaks for Au (111) and for TIG films indicating a good crystallinity of each layer. The inset shows in details the superimposed peaks for the two TIG layers. (c) SQUID magnetometry for TIG(30 nm)/Au(8 nm)/TIG(15 nm) as a function of $H_X$ and $H_Z$. $M$ vs $H_Z$ loop shows independent switches and a PMA for each TIG layer. The in-plane magnetic signal is smaller and higher fields are required to saturate the film. The arrows indicate the magnetization direction of each TIG layer as $H_Z$ changes from positive to negative values. (d) show polar MOKE data for TIG(60 nm)/Au(8 nm)/TIG(30 nm). (e) and (f) Show XRD scan and SQUID magnetometry for TIG(30 nm)/Au(12 nm)/TIG(15 nm). The inset in (e) shows the details around TIG peaks. (g) Polar MOKE data for TIG(30 nm) in black curve, TIG(30 nm)/Au(12 nm) in green, and red curve for TIG(30 nm)/Au(12 nm)/TIG(15 nm). SQUID or MOKE don't show well-separated magnetic switching for each TIG layers. (h) Maintaining 12 nm-thick Au layer and decreasing the thickness of the TIG layers, which strengthens the PMA, MOKE data shows independent switches for each magnetic layer in TIG(15 nm)/Au(12 nm)/TIG(7.5 nm).

The coupling between the TIG layers shows a dependence on the thickness of Au and TIG films. For 8 nm thick Au layer, we observe independent out-of-plane switches for TIG(30 nm)/Au(8 nm)/TIG(15 nm), as shown Fig. 7(c). The out-of-plane magnetization reversal process presents two steps at reversal magnetic fields of 35 Oe, corresponding to the bottom 30 nm thick TIG layer, and – 65 Oe corresponding to the top 15 nm thick TIG layer. When the TIG layers become thicker, while maintaining 8 nm of Au, $\varepsilon_\parallel$ decreases and weakens the PMA. Hence, the polar MOKE for TIG(60 nm)/Au(8 nm)/TIG(30 nm) shows an initial abrupt transition at 170 Oe followed by a continuous reversal process of the TIG layers, as shown in Fig. 7(d). Polar MOKE for TIG/Au/TIG is not as accurate as SQUID because of limitations due to the light penetration depth which is governed mainly by Au layer, since TIG is transparent. It is challenging to access TIG bottom film as Au layer becomes thicker.

Figures 7(e), (f), (g) and (h) show polar MOKE, SQUID, and XRD scans for trilayers with 12 nm thick Au films. The XRD scan in Fig. 7(e) shows peaks for Au(12 nm), TIG(15 nm) and TIG(30 nm). The strain for 30 nm and 15 nm of TIG films are + 0.191 ± 0.005 % and + 0.476 ± 0.005 %, respectively. These values are comparable with the strain for the trilayer with 8 nm of Au. The SQUID and polar MOKE measurements, shown in Fig. 7(f), do not show independent switching for each TIG layer, but on the other hand, they show a more robust PMA with a coercive field for the whole structure of 75 Oe. Figure 7(g) shows a comparison between the simple 30 nm thick TIG film with and without a 12 nm layer of Au, and the trilayer. Adding a top layer of TIG strongly affects the magnetic anisotropy of the system. For a thinner trilayer, TIG(15 nm)/Au(12 nm)/TIG(7.5 nm), the polar MOKE shows independent switchings of the two TIG layers with coercive fields of 6 Oe and 14 Oe, respectively, as can be seen in Fig. 7(h), and the MOKE signal becomes square indicating a robust PMA for both TIG layers.

Our results provide ways to tune the PMA or IPMA in simple films and in more complex multilayers composed of TIG and HMs, by adjusting film thickness, substrate lattice constant and annealing conditions. We show that it is possible to control the magnetic anisotropy in TIG/Au/TIG heterostructures and build a robust PMA in these systems for applications in spintronic devices. Another use is for magnon valves, where it is necessary to build an IPMA on these structures, which could be achieved by lowering the annealing temperature to 600℃ and increasing the thickness of TIG layers.

**4. Conclusion**

In summary, we demonstrate control of the strain-dependent magnetic anisotropy in sputtered ultrathin TIG, by optimally selecting the thickness, post-growth annealing conditions, and substrate. As-grown sputtered TIG films are nearly amorphous (absence of diffraction peaks) and not magnetic, irrespective of different sputtering conditions. Films grown at room temperatures and with post-growth thermal treatment are very smooth and epitaxial, and show a saturation magnetization close to the bulk value for TIG, 110 emu/cm$^3$. 30 nm thick TIG films annealed at 600 °C have an IPMA, and as the annealing temperature increases the perpendicular magnetization saturation and the in-plane strain enhaces favoring a PMA. Film thickness also influences the magnetic anisotropy: thinner films show higher positive $\varepsilon_\parallel$, favoring a robust PMA.

All the films annealed in different atmospheres (oxygen, air, argon, 95% argon + 5% oxygen) are epitaxial with high-quality surface and yield similar magnetic signals, with the exception of vacuum annealed ones, which lack a magnetic signal and crystalline structure. Based on our observations we speculate that the main role of the gas during the annealing process is to prevent the loss of oxygen from TIG film: if so, the gas pressure needs to be high enough to guarantee no film deterioration. An interesting result is that air-annealing is very effective and thus could be a low-cost alternative to crystallize amorphous RIG thin films. It is possible to shorten the duration of annealing without a major reduction in the magnetic signal. We also observe that small changes in the substrate lattice constant lead to dramatic changes in the magnetic anisotropy. Films are found to be magnetically much softer when grown on GGG as compared to those grown on SGGG.

Finally, we demonstrate that it is possible to fabricate epitaxial TIG/Au/TIG trilayers with controllable PMA or IPMA, in which the magnetic coupling between the two TIG layers depends on the thickness of the Au layer. These results open up a range of possibilities for tuning the magnetic anisotropy of TIG films, and therefore leading to implementation of these films in more complex structures and devices. Moreover, having smooth and controlled anisotropy gives the versatility to open a vast area by combining TIG films with other quantum materials such as TIs and SCs to create all *in situ* hybrid structure for exploring new physics through the interfacial exchange interaction [12, 14, 15, 41].

**Acknowledgements**

This research is supported by Army Research Office (ARO W911NF-19-2-0041 and W911NF-19-2-0015), NSF (DMR 1700137), ONR (N00014-16-1-2657), and Brazilian agencies CAPES (Gilvania Vilela/POS-DOC-88881.120327/2016-01), FACEPE (APQ-0565-1.05/14), CNPq and UPE (PFA/PROGRAD/UPE 04/2017). D. S. and Y.O. thanks the Center for Integrated Quantum Materials - NSF (DMR-1231319) for financial support. P. Z. was partly funded by Summer Undergraduate Research Fellowship (SURF) from Caltech during his summer internship at MIT.